\documentclass[twocolumn,prl,aps,superscriptaddress,showpacs]{revtex4}
\usepackage[dvips]{graphicx}
\usepackage{latexsym}
\usepackage{amsmath}
\usepackage{amsfonts}
\usepackage{amssymb}
\usepackage{bm}
\begin{document}
\newcommand{\fig}[2]{\includegraphics[width=#1]{#2}}
\newcommand{\la}{{\langle}}
\newcommand{\ra}{{\rangle}}
\newcommand{\dg}{{\dagger}}
\newcommand{\upa}{{\uparrow}}
\newcommand{\dna}{{\downarrow}}
\newcommand{\ab}{{\alpha\beta}}
\newcommand{\ias}{{i\alpha\sigma}}
\newcommand{\ibs}{{i\beta\sigma}}
\newcommand{\hH}{\hat{H}}
\newcommand{\hn}{\hat{n}}
\newcommand{\hc}{{\hat{\chi}}}
\newcommand{\hU}{{\hat{U}}}
\newcommand{\hV}{{\hat{V}}}
\newcommand{\br}{{\bf r}}
\newcommand{\bk}{{{\bf k}}}
\newcommand{\bq}{{{\bf q}}}
\def\gsim{~\rlap{$>$}{\lower 1.0ex\hbox{$\sim$}}}
\setlength{\unitlength}{1mm}

\title{Extended Hubbard model of superconductivity driven by charge fluctuations
in iron-pnictides}
\author{Sen Zhou}
\affiliation{Department of Physics, Boston College, Chestnut
Hill, Massachusetts 02467, USA} \affiliation{Department of
Physics and Astronomy, Rutgers University, Piscataway, New Jersey
08854, USA}
\author{G. Kotliar}
\affiliation{Department of
Physics and Astronomy, Rutgers University, Piscataway, New Jersey
08854, USA}
\author{Ziqiang Wang}
\affiliation{Department of Physics, Boston College, Chestnut
Hill, Massachusetts 02467, USA}
\date{\today}
\begin{abstract}

We present a scenario for iron-pnictide superconductivity mediated by charge fluctuations that are strongly enhanced by Fe-As intersite electronic interactions. Deriving an eight-band extended Hubbard model including Fe 3$d$ and As 4$p$ orbitals for the LaOFeAs family, we show that charge fluctuations induced by $p$-$d$ charge transfer and As orbital polarization interactions in the Fe-pnictogen structure peak at wavevectors $(0, 0)$, and ($\pi$, 0) and ($\pi$, $\pi$) respectively. Intraorbital spin-singlet pairing attraction develops at these wavevectors and the solution of the linearized gap equation shows robust $s$-wave superconductivity with both $s_\pm$ and $s_{++}$ gap functions.

\typeout{polishabstract}
\end{abstract}
\maketitle

The mechanism of high-T$_c$ superconductivity in the Fe-pnictides has attracted enormous attention since its original discovery in F-doped LaFeAsO (1111) \cite{sc1111}. The majority of the theoretical efforts has focused on the proximity of the superconducting (SC) phase to the spin density wave (SDW) state and the multiple Fermi surfaces (FS) associated with the Fe $3d$ and As $4p$ orbitals \cite{mazin08,kuroki08,jphu08,fwang09,graser09,yanagi}. An emerging picture is that spin fluctuations and FS scattering favor spin-singlet $s_\pm$-wave pairing where the gap function changes sign between hole and electron FS due to the intraorbital repulsion in the particle-particle channel. For the prototypical 1111 series, where $T_c$ reaches the record high of 55K when La is replaced by other rare earths
\cite{highertc}, NMR Knight shift measurements indeed find spin-singlet pairing \cite{nmr2},
but it remains unclear whether spin fluctuations are the driving force for superconductivity. Upon electron-doping, spin fluctuations in the normal state are dramatically suppressed; the SDW phase terminates abruptly and is separated from the SC state by a first order-like transition \cite{Nakai,Grafe,Mukuda,Luetkens}. The correlation between $T_c$ and the low energy spin-fluctuations measured by the spin-lattice relaxation rate has been found to be rather weak \cite{Nakai,Grafe,Mukuda}. Moreover, applying pressure near the optimal doping level increases $T_c$ from 23K to 43K while the strength of spin-fluctuations remains unchanged \cite{Nakano2010}. This is further supported by recent muon spin rotation ($\mu$SR) and magnetization experiments in the underdoped regime that show hydrostatic pressure suppresses magnetic interactions but strongly enhances $T_c$ \cite{muon}. Thus, spin-fluctuations alone cannot fully account for the pairing mechanism of iron-pnictide superconductors.

In this paper, we explore a different scenario where the superconductivity is driven by charge fluctuations.
There are indeed emerging experimental evidence that the pnictides are close to the charge ordering instability. In the 1111 series, two distinct charge environments are detected by As NQR measurements in the underdoped regime, indicative of local electronic charge order \cite{NQR,Zheng}.
In contrast to the cuprates, the Fe-pnictides are $p$-$d$ charge transfer metals with low energy charge fluctuations. It is thus important to go beyond the local Hubbard interactions and consider the interatomic interactions. Furthermore, due to the large spatial extent of the As 4$p$ orbital, the interactions between the Fe 3$d$ and As 4$p$ electrons are important both in the charge transfer channel and in the As orbital polarization channel when charges fluctuate at the Fe site. We found that it is a generic feature of the Fe-pnictogen structure that these interactions produce enhanced charge fluctuations at $(0, 0)$, $(\pi, \pi)$, and $(\pi,0)$ respectively,
and mediate attractions for intraorbital pairing at these wavevectors.

We focus on the electron-doped 1111 series that shares a single Fe-pnictogen layer per unit cell and is the most quasi-two-dimensional Fe-pnictides. We derive an extended Hubbard model as the low energy effective Hamiltonian for the FeAs layer: $\hH=\hH_0+\hU_{dd}+\hV_{pd}$, where $\hH_0$ is a tight-binding model for the band structure including both the Fe 3$d$ and As 4$p$ orbitals; $\hU_{dd}$ describes the local interactions, intraorbital Hubbard repulsion $U$ and Hund's rule coupling $J$, at the Fe sites; and $\hV_{pd}$ contains the nearest neighbor (NN) charge transfer interaction $V$ and As orbital polarization interactions $\Delta V_1$ for $p_x$-$p_y$ and $\Delta V_2$ for $p_z$-$p_{x,y}$. Treating $U$ and $J$ as effective interaction parameters, a random phase approximation (RPA) study of the charge and spin fluctuations is carried out as a function of the Fe-As inter-site interactions. We find that the enhanced charge fluctuations lead to robust $s$-wave superconductivity with both $s_\pm$ and $s_{++}$ gap symmetry as summarized in Table I for a wide range of doping levels.
\begin{table}[htb]
\caption{\label{symmetry} Symmetry of the leading pairing instability driven by $p$-$d$ interactions for different on-site $U$ and Hund's rule coupling $J$. All cases listed are nodeless.}
\begin{ruledtabular}
\begin{tabular}{cc|ccc}
$J/U$ & $U({\rm eV})$ & $V$-driven & $\Delta V_1$-driven & $\Delta V_2$-driven \\
\hline
0.1 & 0.6 & $s_{++}$ & $s_\pm$ & $s_{++}$ \\
0.1 & 1.2 & $s_\pm$ & $s_\pm$ & $s_{++}$ \\
\hline
0.3 & 0.5 & $s_{++}$ & $s_\pm$ & $s_{++}$ \\
0.3 & 1 & $s_\pm$ & $s_\pm$ & $s_{++}$ \\
\end{tabular}
\end{ruledtabular}
\end{table}

The low energy part of the La1111 band dispersions shown in Fig.~1a can be described by a tight-binding model $H_0$ for the Fe 3$d$ and As 4$p$ complex \cite{yanagi}. For the single-layered 1111, it is possible to unfold the reduced zone to the original one corresponding to one FeAs per unit cell and work with 8 bands specified by an orbital index $a=$ 1($d_{xy}$), 2($d_{yz}$), 3($d_{zx}$), 4($d_{x^2-y^2}$), 5($d_{3z^2-r^2}$), 6($p_x$), 7($p_y$), 8($p_z$). Fig. 1a shows that the $p$-$d$ model $H_0$ describes well both the LDA band dispersion and the orbital character for the undoped case with 12 electrons per unit cell. At $10\%$ electron doping, the FS contain two hole pockets (labeled by $\alpha$ and $\beta$) centered around $\Gamma$ and two electron pockets around X (labeled by $\gamma$) and Y pionts. Fig.~1b and 1c display the dominant Fe 3$d$ and As 4$p$ orbital characters on the FS respectively.
\begin{figure}
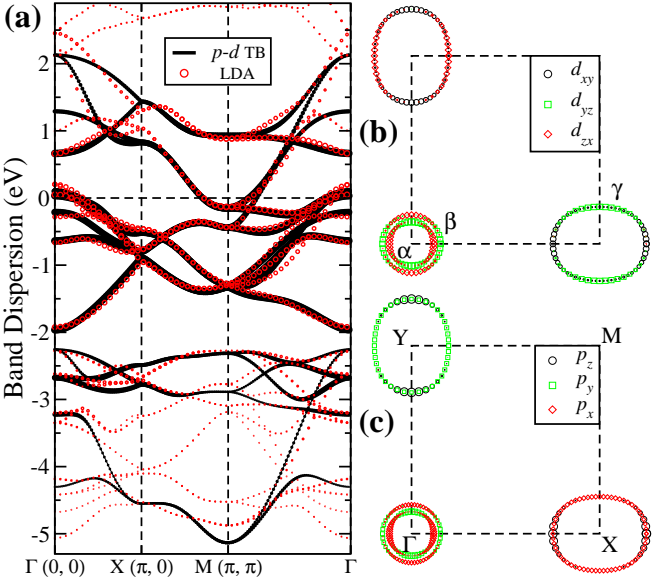

\begin{center}
\fig{3.4in}{fig1.eps}
\caption{(Color online){\it Eight-band $p$-$d$ model}
(a) Comparison of the band dispersions to LDA band structure in the reduced zone.
Line thickness and symbol size denote Fe 3$d$ content. Fe 3$d$ (b) and As 4$p$ (c) contributions to the FS in the unfolded zone at 10\% electron doping. Symbol sizes denote the orbital content with those of the As 4$p$ enhanced by a factor of 4.}
\label{fig1}
\end{center}
\vskip-7mm
\end{figure}

The electronic interactions have the general form
\begin{equation}
\hH_I = {1\over 2}\sum_{ij,\sigma\sigma'}\sum_{ab,a'b'} W_{ab,a'b'}
(\br_{ij}) c^\dagger_{ia\sigma} c^\dagger_{jb'\sigma'}
c_{ja'\sigma'} c_{ib\sigma}
\end{equation}
where $c^\dagger_{i a \sigma}$ creates a spin-$\sigma$ electron on
site $i$ in orbital $a$. The Coulomb integral is given by
\begin{equation}
W_{ab,a'b'}(\br_{ij}) = \int d^3\br d^3\br' \phi^*_{a}(\br)
\phi^*_{b'}(\br') V(R_{ij}) \phi_{a'}(\br') \phi_b(\br),
\label{coulombint}
\end{equation}
where $R_{ij}=\vert\br_{ij}+\br'-\br\vert$ and $\phi_a$ is the wavefunction of orbital $a$. Retaining the dominant on-site interactions for the Fe atoms (those of the As are much weaker) and the NN $p$-$d$ interactions, we write
$\hH_I=\hU_{dd}+\hV_{pd}$. $\hU_{dd}$ attains the usual multi-orbital Hubbard model
\begin{align}
\hU_{dd}&=U\sum_{i,\alpha}\hat{n}_{i\alpha\upa} \hat{n}_{i\alpha\dna}
+\left(U'-{1\over 2}J\right)
\sum_{i,\alpha<\beta} \hat{n}_{i\alpha} \hat{n}_{i\beta} \label{h1} \\
&-J\sum_{i,\alpha\neq\beta}{\bf S}_{i\alpha}\cdot {\bf S}_{i\beta}
+J'\sum_{i,\alpha\neq\beta} c^\dg_{i\alpha\upa} c^\dg_{i\alpha\dna}
c_{i\beta\dna}c_{i\beta\upa}, \nonumber
\end{align}
with intra and inter orbital on-site Coulomb repulsions $U= W_{\alpha\alpha, \alpha \alpha}
(0)$, $U'= W_{\alpha \alpha,\beta\beta}(0)$ and the Hund's rule coupling $J= J'= W_{\alpha
\beta, \alpha \beta}(0)$. Orbital rotation symmetry requires $U=U'+2J$.
%
Here and henceforth, we use $\alpha,\beta=1,2,\cdots,5$ to distinguish Fe 3$d$ orbitals from  As 4$p$ orbitals denoted by $\mu,\nu=6,7,8$.

The Coulomb integral $W_{\alpha\beta,\mu\nu}$ describes a rich variety of Fe-As interatomic interactions. The $p$-$d$ charge transfer interaction
$V_{\alpha,\mu}=W_{\alpha\alpha,\mu\mu}(\br^*)$ where $\br^*$ is the vector connecting the NN Fe and As.
The importance of $V$ was emphasized in the context of the cuprate superconductivity \cite{varma89}. Furthermore,
$\Delta V_{\alpha,\mu\nu}=W_{\alpha\alpha, \mu\nu}(\br^*) $ describes the As 4$p$ orbital polarization induced by the Fe electric field associated with the charge fluctuations in the $\alpha$-orbital.
This is different from the higher energy As 4$p$-5$s$ polarizations discussed in Ref.~\cite{sawatzky,bishop}. The large spatial extent of the As 4p orbital has important consequences: (i) The bare interaction $\Delta V_{1,2}$ estimated using the hydrogen-like atomic wavefunctions in Eq.~(\ref{coulombint}) is remarkably large and about 10-20\% of the $p$-$d$ charge transfer $V$. Since $V$ is subject to charge screening whereas $\Delta V_{1,2}$ is not, the effective interaction strengths can be comparable. (ii) The interaction involving the polarization of the smaller Fe orbitals $W_{\ab,\mu\mu}$ and the interaction between the Fe and As polarization clouds $W_{\ab, \mu\nu}$ are at least one or two orders of magnitude smaller and can thus be neglected. (iii) Since the 3$d$ orbitals are much smaller, their dependence in $V$ and $\Delta V$ can be ignored. We thus arrive at the following Hamiltonian for the $p$-$d$ interactions,
\begin{align}
\hV_{pd}&=V\sum_{\langle i,j\rangle} \hn_{i}^d\hn_j^p +
\Delta V_1\sum_{\langle i,j\rangle,\sigma}\tau_{ij}^{xy}\hn_i^d\left(p_{x,j\sigma}^\dagger p_{y,j\sigma} + h.c.\right)\nonumber \\
&+\Delta V_2\sum_{\langle i,j\rangle,\sigma}\tau_{ij}^{x(y)z}\hn_i^d\left[p_{z,j\sigma}^\dagger p_{x(y),j\sigma}  + h.c.\right], \label{pdrealspace}
\end{align}
where ${\hat n}_i^d$ and ${\hat n}_j^p$ are the total density operators of the $d$ and $p$ electrons respectively. Since the FeAs block deviates from the ideal tetrahedron structure, two interaction parameters, $\Delta V_1$ and $\Delta V_2$, are introduced to distinguish between As $p_x$-$p_y$ and $p_z$-$p_{x,y}$ orbital polarizations. Note that the polarization (quadrupole) term is orientation-dependent and  $\tau_{ij}^{\mu\nu}$ accounts for the sign of the wavefunction overlap. In momentum space, the $p$-$d$ interaction reads
\begin{equation}
\hV_{pd}= \sum_{\bq\bk} \sum_{\mu\nu,\sigma}
F_{\mu\nu}(\bq) \hn^d(q)
c^\dagger_{\bk+\bq,\nu\sigma} c_{\bk\mu\sigma},
\label{pd-momentum}
\end{equation}
where the form factors $F_{\mu\mu}(\bq)$ $=4V$ $\cos{{1\over 2}q_x}$
$\cos{{1\over 2}q_y}$, $F_{67}(\bq)$ $=-4\Delta V_1$ $\sin{{1\over
2}q_x}$ $\sin{{1\over 2}q_y}$, $F_{68}(\bq)$ $=-i4\Delta V_2$
$\sin{{1\over 2}q_x}$ $\cos{{1\over 2}q_y}$, and $F_{78}(\bq)$
$=-i4\Delta V_2$ $\cos{{1\over 2}q_x}$ $\sin{{1\over 2}q_y}$.

We next present a complete RPA treatment of the interactions in Eqs.(\ref{h1}) and (\ref{pdrealspace}). The charge and spin susceptibilities can be written as $34\times34$ matrices
\begin{align}
\hc^s(\bq,\omega_l)& =\hc^0(\bq,\omega_l)/ [1-\hU^s\hc^0 (\bq,\omega_l)] , \label{rpa}\\
\hc^c(\bq,\omega_l)& =\hc^0(\bq,\omega_l)/ [1+ (\hU^c +2\hV^c(\bq))
\hc^0(\bq,\omega_l) ] \nonumber
\end{align}
where the bare susceptibilities $\chi^0_{ab,a'b'}(\bq,\omega_l)= -{(T/ N)} \sum_{\bk,m} G^0_{aa'}(\bk+\bq,\epsilon_m+\omega_l) G^0_{b'b} (\bk,\epsilon_m)$ with the noninteracting Green's function $\hat{G}^0(\bk,\epsilon_m)= [i\epsilon_m-\hH_0(\bk)]^{-1}$.
In Eq.~(\ref{rpa}),
the nonzero elements of the interaction matrices $\hU^s$, $\hU^c$, and $\hV^c$ are:
$U^s_{\alpha\alpha, \alpha\alpha} =U$,
$U^s_{\ab, \ab} =U'$, $U^s_{\alpha\alpha, \beta\beta} =J$,
$U^s_{\ab, \beta\alpha} =J'$, $U^c_{\alpha\alpha, \alpha\alpha} =U$,
$U^c_{\ab, \ab} =2J-U'$, $U^c_{\alpha\alpha, \beta\beta} =2U'-J$,
$U^c_{\ab, \beta\alpha} =J'$, and $V^c_{\alpha\alpha, \mu\nu} (\bq)
=F_{\mu\nu}(\bq)$.
The on-site interaction enhances (reduces) the spin (charge) susceptibility. The inter-site $p$-$d$ interaction ${\hat V}^c$, on the other hand, affects only the charge sector, entering $\hc^c$ in the block-off-diagonal elements in the denominator. They lead to enhanced charge fluctuations at wavevectors where the interactions $F_{\mu\nu}(q)$ in Eq.~(\ref{pd-momentum}) are maximum in momentum space, i.e., at ${\bf Q}=(0,0)$ for $V$; $(\pi,\pi)$ for $\Delta V_1$; $(\pi,0)$ and $(0,\pi)$ for $\Delta V_2$.

We shall describe our results for $10\%$ electron doping with a moderate effective $U=1$eV, but a reasonably large ratio $J/U=0.3$ in accord with the large Hund's rule coupling in the pnictides (last row in Table I).
Several prominent intraorbital static charge susceptibilities $\chi^c_{\alpha\alpha,\alpha\alpha}(\bq)$ are shown for $V$=0.26 eV (Fig.~2a), $\Delta V_1$=0.3 eV (Fig.~3a), and $\Delta V_2$=0.28 eV (Fig.~4a), independently. Clearly, the inter-site interactions enhance the intraorbital charge fluctuations by introducing peaks at the corresponding ${\bf Q}$ that grow with increasing $V$ and $\Delta V_{1,2}$. We verified that their emergence is tied to the softening of the collective modes in the imaginary part of the dynamical charge and charge transfer susceptibility \cite{tobepublished}. Note that the $p$-$d$ interactions in Eq.~(\ref{pd-momentum}) leave the Fe 3$d$ interorbital susceptibility $\chi^c_{\alpha\beta,\beta\alpha}$ unchanged.
\begin{figure}
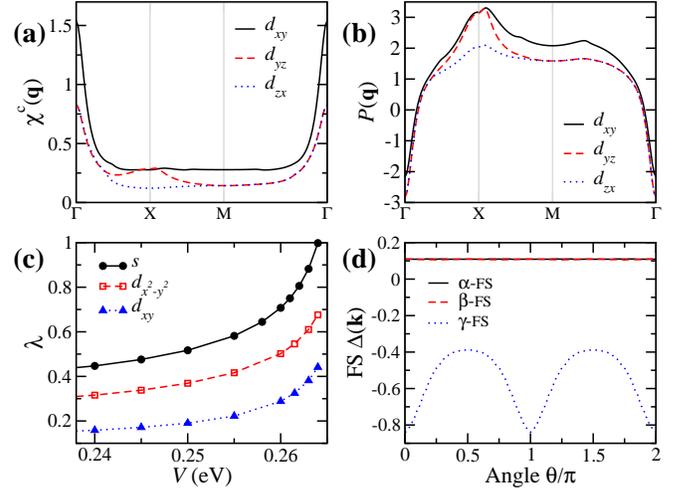

\begin{center}
\fig{3.4in}{fig2.eps} \caption{(Color online).
{\it Effects of $p$-$d$ charge transfer $V$} at $(U, J)$=(1, 0.3) eV. (a) Intraorbital
RPA charge susceptibility and (b) Singlet intraorbital pairing interaction at $V$=0.26 eV. (c) $s$- and $d$-wave eigenvalues $\lambda$ as a function of $V$. (d) $s$-wave gap symmetry function along three FS sheets at
$V$=0.264 eV where $\lambda_s$=1. Angles are measured from $x$-axis.}
\label{fig2}
\end{center}
\vskip-6mm
\end{figure}

To study superconductivity, we evaluate the pairing vertex dressed by the spin and charge fluctuations \cite{bickers89,takimoto04}. The effective spin-singlet pairing interaction is given by
\begin{align}
\hat{P}(\bq)=&{1\over 2} \hU^s + {3\over 2} \hU^s
\hc^s(\bq) \hU^s +{1\over 2}  [\hU^c +2\hV^c(\bq)] \nonumber \\
-&{1\over 2} [\hU^c +2\hV^c(\bq)] \hc^c(\bq) [\hU^c +2\hV^c(\bq)],
\end{align}
where $\hc^{s,c}(\bq)=\hc^{s,c}(\bq,\omega_l=0)$ are the static spin and charge susceptibilities. The spin-triplet pairing turns out to be sub-leading.
The calculated $\hat{P}(\bq)$
are shown in Figs.~2b, 3b, and 4b for interactions $V$, $\Delta V_1$, and $\Delta V_2$ respectively. Remarkably, with the enhancement of the charge fluctuations near ${\bf Q}$ (peaks), the repulsion is weakened (dips) in the intraorbital pairing potential ${P}_{\alpha\alpha,\alpha\alpha}$ and turns into attraction for intraorbital pairing near ${\bf Q}$ when the corresponding $p$-$d$ interaction is sufficiently strong. This is in contrast to the pairing interactions mediated by spin fluctuations that are repulsive at all $\bq$. The SC instability can be obtained by solving the linearized gap equation,
\begin{align}
\lambda \Delta_{ab}(\bk) =&-{T\over N}\sum_{\bk',n} \sum_{a'b',a''b''}
P_{aa'',b''b} (\bk-\bk') \label{gapequation} \\
&\times G^0_{a''a'}(\bk',\omega_n) G^0_{b''b'}(-\bk',-\omega_n)  \Delta_{a'b'} (\bk')
\nonumber
\end{align}
in the orbital basis, where $\Delta_{ab}(\bk)$ is an $8\times 8$ normalized gap symmetry function. The pairing instability sets in when the largest eigenvalue $\lambda$ reaches unity at $T=T_c$. To overcome the finite-size effects, we solved Eq.~(\ref{gapequation}) self-consistently at $T$= 20 meV on an $80\times80$ momentum mesh to obtain $\lambda$ and $\Delta_{ab}(\bk)$ as a function of $V$ and $\Delta V_{1,2}$. The gap symmetry function can be easily transformed into the band basis by a unitary rotation and plotted along the FS.

\begin{figure}
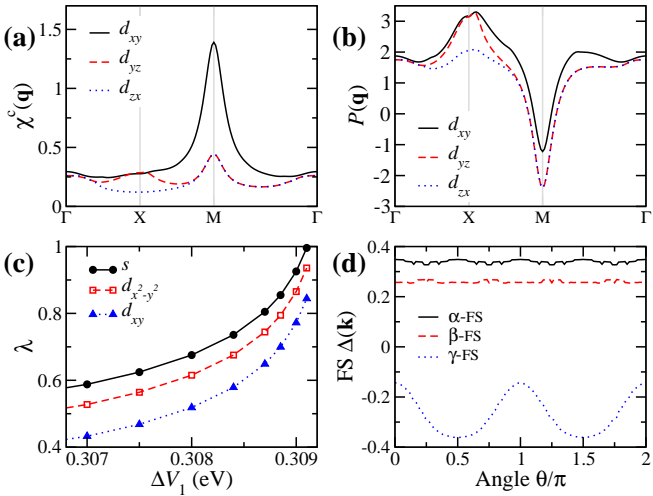

\begin{center}
\fig{3.4in}{fig3.eps} \caption{(Color online)
{\it Effects of $p_x$-$p_y$ orbital fluctuation $\Delta V_1$} at $(U, J)$=(1, 0.3) eV. (a) Intraorbital
RPA charge susceptibility and (b) Singlet intraorbital pairing interaction at $\Delta V_1$=0.3 eV. (c) $s$- and $d$-wave eigenvalues $\lambda$ as a function of $\Delta V_1$. (d) $s$-wave gap symmetry function along three FS sheets at
$V$=0.309 eV where $\lambda_s$=1. Angles are measured from $x$-axis.}
\label{fig3}
\end{center}
\vskip-6mm
\end{figure}

{\it Superconductivity driven by inter-site interaction $V$} is summarized in Fig.~2. The eigenvalues $\lambda$ plotted as a function of $V$ in Fig.~2c show that $s$-wave pairing is more favorable than pairing with $d$-wave symmetries and superconductivity sets in at a reasonably small $V_c=0.264$eV. The normalized gap symmetry function in Fig.~2d shows that the pairing symmetry is the nodeless $s_\pm$-wave; with opposite signs for the pairing gaps on the electron $(\gamma)$ and the hole ($\alpha$ and $\beta$) pockets. The obtained $\Delta_{ab}(\bk)$ in the orbital basis shows that all orbitals, including those of the As 4$p$, contribute in a complicated manner to the behavior of the gap function on the FS. Nevertheless, the pairing symmetry can be qualitatively understood from the dominant intraorbital pairing interactions shown in Fig.~2b. While the increasing attraction peaked around $(0,0)$ provides the main pairing force through forward scattering in contrast to spin fluctuation mediated pairing, the scattering by the repulsion near $(\pi,0)$ and $(0,\pi)$ favors a sign change between the electron and the hole pockets in a similar manner as in the spin fluctuation scenario \cite{mazin08,kuroki08,yanagi}. Furthermore, the repulsion near $(\pi,\pi)$ causes a degree of frustration for the $s_{\pm}$-pairing, leading to the large asymmetry of the gap function and large variations on the electron FS. Remarkably, keeping the same ratio $J/U=0.3$, but reducing the Hubbard $U$ by a factor of two, we find that the pairing symmetry changes to the $s_{++}$-wave due to the reduction in the repulsion at finite momenta associated with spin-fluctuations. The change from $s_{\pm}$ pairing at large $U$ to $s_{++}$ pairing at small $U$ is also true for a smaller ratio of $J/U=0.1$ and may be generic of the SC phase driven by the $p$-$d$ charge transfer interaction $V$ (Table I).

\begin{figure}
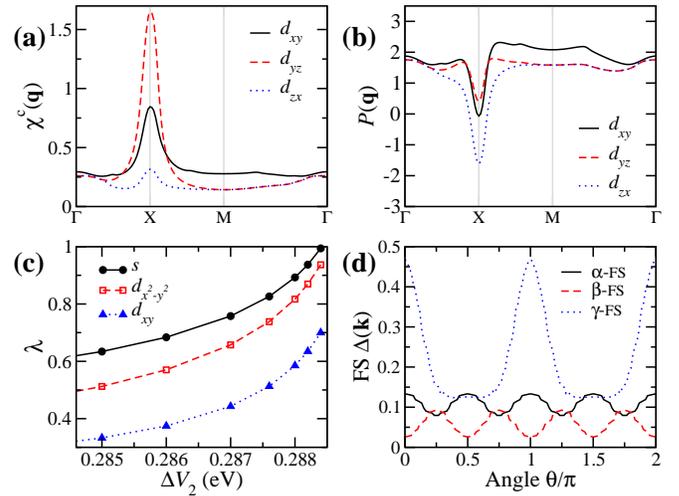

\begin{center}
\fig{3.4in}{fig4.eps} \caption{(Color online).
{\it Effects of $p_z$-$p_{x,y}$ orbital polarization $\Delta V_2$} at $(U, J)$=(1, 0.3) eV. (a) Intraorbital
RPA charge susceptibility and (b) Singlet intraorbital pairing interaction at $\Delta V_2$=0.28 eV. (c) $s$- and $d$-wave eigenvalues $\lambda$ as a function of $\Delta V_2$. (d) $s$-wave gap symmetry function along three FS sheets at
$\Delta V_2$=0.288 eV where $\lambda_s$=1. Angles are measured from $x$-axis.}
\label{fig4}
\end{center}
\vskip-6mm
\end{figure}

{\it Superconductivity driven by inter-site interaction $\Delta V_{1}$} is summarized in Fig.~3. The largest eigenvalues of the gap equation plotted in Fig.~3c show that $s$-wave pairing dominates over $d$-wave symmetries and the SC phase sets in at $\Delta V_{1,c}=0.309$eV. The gap symmetry function over the FS shown in Fig.~3d reveals that the pairing symmetry is the sign-changing $s_\pm$-wave. Remarkably, the gap over the electron pocket oscillates moderately around a value that is close in magnitude to that on the inner hole pocket, but larger than that on the outer hole pocket, in excellent agreement with the gap ratios observed by ARPES in optimally doped K$_{x}$Ba$_{1-x}$Fe$_2$As$_2$ \cite{dingEPL}. Moreover, we find that the nodeless $s_{\pm}$ pairing symmetry is a robust feature of the superconductivity driven by Fe charge fluctuations coupled to As $p_x$-$p_y$ orbital polarization for different values of $U$ and $J/U$ as shown in Table I. This remarkable feature is a result of the pairing interaction shown in Fig.~3b. The repulsion at $(\pi,\pi)$ has been turned into the growing attraction by $\Delta V_1$ that provides the main pairing force through $(\pi,\pi)$-scattering, leaving the repulsion at $(\pi,0)$ and $(0,\pi)$ unfrustrated that locks the opposite sign of the gap functions on the electron and hole pockets.

{\it Superconductivity driven by inter-site interaction $\Delta V_{2}$} is summarized in Fig.~4. It is clear from Fig.~4c that the leading SC instability remains in the s-wave channel and sets in at $\Delta V_{2,c}=0.288$eV. The pairing interaction in Fig.~4b shows that $\Delta V_2$ has turned the repulsion at $(\pi,0)$ and $(0,\pi)$ due to primarily spin-fluctuations into the growing attraction which serves as the dominate pairing force in this case. As a result, the $s_{\pm}$ symmetry becomes unfavorable. Indeed, the gap symmetry function shown in Fig.~4d reveals an anisotropic $s_{++}$-wave with significant variations on the electron pocket. We find that the $s_{++}$-wave pairing is a robust feature of the superconductivity driven by $\Delta V_2$ for different values of $U$ and $J/U$, as shown in Table I.

In summary, we proposed that the iron-pnictides superconductivity can be driven by charge fluctuations.
The inter-site interactions in the Fe-pnictogen structure are found to produce strong charge fluctuations that mediate attractions in the spin-singlet pairing potential around wavevectors $(0, 0)$, $(\pi, \pi)$, and $(\pi,0)$. For electron doped LaFeAsO, moderate Fe-As intersite interaction strengths can induce superconductivity with robust $s$-wave symmetry; both sign-changing $s_{\pm}$ and sign-preserving $s_{++}$ gap functions are possible. We suspect that electron-phonon coupling \cite{Kontani} may play a role in such a pairing mechanism, particularly because these wavevectors are the same as the possible lattice instability vectors. It is also tempting to speculate that the $1\times2$ and $\sqrt{2}\times\sqrt{2}$ structures observed by STM in (Ba,Sr)Fe$_2$As$_2$ \cite{Nascimento09} are related to the strong As orbital fluctuations in the bulk pinned by the surface potential. The strong charge fluctuations can be pinned by impurities and defects in the bulk of the sample, leading to local charge order and/or orbital polarization that should be observable to local probes such as NMR and $\mu$SR and serve as a test of the present theory through their correlations with the SC transition temperature.

This work is supported in part by DOE DE-SC0002554, DE-FG02-99ER45747, and NSF DMR-0906943. We thank Y. Yanagi, Y. Yamakawa, H. Ding, V. Madhavan, and S.-H. Pan for useful discussions.

\end{document}